\documentclass[12pt,onepage,oneside]{article}
\usepackage{fullpage}
\usepackage{amsfonts,amsmath,amssymb}
\usepackage{graphicx}
\usepackage{dcolumn}
\usepackage{siunitx}
\usepackage{tabularx}
\usepackage{booktabs}
\usepackage{xcolor,colortbl}
\usepackage{color}
\usepackage{comment}
\usepackage{setspace}
\usepackage{bbm}
\usepackage{natbib}
\usepackage{threeparttable}
\usepackage{multirow}
\usepackage{lscape}
\usepackage{pdflscape}
\usepackage{rotating}
\usepackage{graphicx}
\usepackage{times}
\usepackage{float}
\usepackage[titletoc,toc,title]{appendix}
\usepackage{subcaption}
\usepackage{color}
\newcommand{\vix}{$\mathrm{IVIX}$ }
\newcommand{\vixm}{$\mathrm{IVIX^-}$ }
\newcommand{\vixp}{$\mathrm{IVIX^+}$ }

\newcommand{\oa}{Online Appendix~}

\def\Gray{\color{lightgray}}

\usepackage{amsopn}
\usepackage[colorlinks=true,linkcolor=black,allcolors=black]{hyperref}
\graphicspath{ {LateX} }
\usepackage{pdflscape}
\usepackage{rotating}
\usepackage{float}
\usepackage[titletoc,toc,title]{appendix}
\usepackage{amsmath}
\usepackage{authblk}
\usepackage{subcaption}
\graphicspath{ {LateX} }
\usepackage[left=1in,right=1in,top=1.in,bottom=1.in]{geometry}
\usepackage{times}

\begin{document}

\title{Asymmetric Network Connectedness of Fears
\footnote{We are grateful to the editor, Bryan Graham, and three anonymous referees for their useful comments and suggestions, which have greatly improved the paper. We are indebted to Jon Danielsson, Robert Faff, Jaideep Oberoi, Sotiris Staikouras, Jean-Pierre Zigrand, and the participants at the Royal Economic Society annual conference 2019, IESD 2019, and CFE 2019 for many useful comments, suggestions, and discussions. Jozef Barunik gratefully acknowledges support from the Czech Science Foundation under the EXPRO GX19-28231X project. Mattia Bevilacqua gratefully acknowledges the support of the Economic and Social Research Council (ESRC) in funding the Systemic Risk Centre [grant number ES/K002309/1 and ES/R009724/1]. A number of additional results accompanying the main analysis are relegated to the \oa published with the paper.}}

\author[1]{Jozef Barun\'{\i}k}
\author[2]{Mattia Bevilacqua}
\author[3]{Radu Tunaru}
\affil[1]{The Czech Academy of Sciences, Institute of Information Theory and Automation; Institute of Economic Studies, Charles University}
\affil[2]{Systemic Risk Centre, London School of Economics}
\affil[3]{ University of Sussex}


\maketitle
\begin{abstract}

This paper introduces forward-looking measures of the network connectedness of fears in the financial system, arising due to the good and bad beliefs of market participants about uncertainty that spreads unequally across a network of banks. We argue that this asymmetric network structure extracted from call and put traded option prices of the main U.S. banks contains valuable information for predicting macroeconomic conditions and economic uncertainty, and it can serve as a tool for forward-looking systemic risk monitoring.

\end{abstract}


\newpage
\section{Introduction}

The financial sector plays an important role in the functioning of the economy through intermediation. Shocks to the financial system impact the real economy largely, yet despite the enormous efforts of researchers and policy makers, we do not understand the mechanism fully. Similar to consumers, firms, and countries creating ever-intensifying linkages in a world economy, the financial sector is connected more than ever before. A critical issue for central bankers and policy makers is how to measure such network connections and understand how they are related to future economic downturns. However, network connectedness remains an incompletely defined concept, and the impact of the financial network structure on the real economy remains poorly understood.

To contribute to this debate, we develop a novel forward-looking set of measures of connectivity in the financial system and study their usefulness in relation to the real economy. Using traded option prices, we measure how market fears stemming from uncertainty\footnote{Note that option prices are often used to measure the forward-looking volatility of the whole market in the financial literature \citep[see][]{fleming1995,christensen1998,Whaley2008}. Moreover, \cite{santa2010} argued that the measures extracted from options are the ex ante risks assessed by option investors. Our definition of fears and their measurement using option prices for individual banks are discussed in detail in section \ref{DataMeth}.} about future price fluctuations covary across financial companies and how shocks to these fears create a network and spread within that network. We argue that the beliefs of investors buying call and put options are linked in a different way and that they can be used to extract asymmetric information about the network structure of the financial system. Exploring these networks of good and bad fears, we show that the information contained in the novel forward-looking measures of network connectedness is valuable for forecasting macroeconomic conditions as well as economic uncertainty measures.
In contrast to the previous literature measuring ex post systemic risk \citep[see][]{billio2012,diebold2014,hautsch2014,hardle2016,geraci2018}, we aim to provide an ex ante systemic risk \emph{alarm bell} that is useful for anticipating the propagation of risk in the financial sector.

Working with fear as a function of the good and bad outcomes expected by option buyers, a key ingredient of our approach is the measurement of such information from traded option prices. While good and bad volatility measures are well established in the literature \citep[see][]{barndorff2010,patton2015,segal2015,feunou2017,kilic2018,Bollerslev2018}, these notions applied to asymmetric responses of volatility to surprise shocks are exclusively based on ex post measures. Our novelty is in proposing forward-looking measures in the network context, as well as studying how shocks to market agents' expectations on both sides of the market create asymmetric linkages in the financial network. With the forward-looking information about the two sides of the financial network, it is useful to study its information content for the development of the real economy, especially about economic downturns. For this purpose, we first construct a new dataset of daily forward-looking volatility measures separately from the traded call and put option prices of major financial institutions, representing the financial network of the U.S. economy.\footnote{Our focus on the network connectedness of the main U.S. financial stocks is motivated by the fact that financial institutions have always been under the magnifying glass of investors, practitioners and academics for their pivotal role in systemic risk terms. Excellent discussions along this line can be found in \cite{billio2012}, \cite{diebold2014}, \cite{barunik2018} and \cite{geraci2018}. The financial sector's systemic risk exposure may lead to macroeconomic decline and macroeconomic contagion \citep{allen2012}, so it should be closely monitored.} Then, we construct the asymmetric network of the two sides of the market expectations to investigate the structural characteristics of the financial system.

The network measures are built in the tradition of dynamic predictive modeling under misspecification, and important causal linkages are approximated via vector autoregression models \citep{diebold2009,Diebold2012,diebold2014}. The connectedness from put option prices measures how shocks to investors' expectations associated with bad volatility, which could be linked to a possible decrease in economic growth \citep[see][]{segal2015,barunik2016,feunou2017,Bollerslev2018}, travel across the network. When constructed from call option prices, the network measure indicates how shocks to a positive direction associated with events that may trigger higher returns \citep{diebold2015} and good volatility spread across the system. Ultimately, we document the in-sample and out-of-sample predictive power of the asymmetric network connectedness with respect to macroeconomic and uncertainty indicators, and we find that they can be related to future economic activities.

\section{Good and Bad Fear, Investor Beliefs and Option Implied Volatility}
\label{DataMeth}

Academics, policy makers, and practitioners fear uncertainty regarding future price fluctuations measured by volatility. To study the network of such fears in the market, we develop forward-looking measures reflecting investor beliefs from data. One of the most popular measures of expectations about uncertainty is the volatility implied by traded option prices. To track investors' beliefs and to allow for one to trade on forward-looking volatility, the Chicago Board Options Exchange introduced a popular volatility index---VIX---extracting expectations from options prices in a model-free manner. The concept was later formalized by \cite{Bakshi&Madan2000,Bakshi2003} and has quickly gained popularity in the literature as well as among practitioners and policy makers as an ``investor fear gauge'' \citep{diebold2014}.

Occurring in response to the anticipation of perceived risks, fear has negative connotations and is often mistaken for bad events or threats. However, fear may be more complicated. More nuanced and referring to a variety of situations, fear can be thought of as a function of an event. A possible positive or negative outcome of an event itself then determines the perception of such fear. As an example, let us consider a possible increase in stock price in response to the announcement of a friendly takeover. Per se, the increased price fluctuation signals increased uncertainty in the market, hence greater risk and fear. The outcome of the transaction is however positive for all the participants, and the increased uncertainty is connected to the positive event in this case. There is always a chance of the expectation not being realized, but the event is the deciding factor.

Since investors tend to react differently to ``state-dependent'' uncertainty and because markets move with positive and negative expectations of investors, it is important to be able to gauge those beliefs. By the use of good fear and bad fear, we label these two complementary situations. Beliefs connected to a good state (bad state) of the economy---good (bad) fears---reflect the situation where an investor fears uncertainty regarding price fluctuations, but the uncertainty itself is connected to a positive (negative) outcome. The realization of the positive outcome, connected to a good fear,  resulting in stock price increases (and therefore firm value) signals strong economic performance. Moreover, the realization of the negative outcome, connected to a bad fear, results in falling prices and therefore in deteriorating firm value, which signals weak economic performance.

To capture the good and bad fears, we use forward-looking uncertainty measures that are intimately related to the VIX methodology inferring fears from traded option prices in a model-free manner. Instead of looking at the whole U.S. stock market index, we are interested in measuring fears at the individual company level, more specifically focusing on the U.S. financial sector. Hence, we develop a methodology to measure uncertainty about individual companies disentangling the aggregate implied volatility into good and bad volatilities connected to the positive and negative state of markets, respectively.

\subsection{Inferring Good and Bad Fear from Option Prices}
\label{vixmeth}

Formalizing the discussion, we consider the price of a volatility contract that pays off the squared log return $R_{t+1}^2 = (p_{t+1} - p_t)^2$ at time $t+1$ with $p_t$ denoting the natural logarithm of the share price $P_t$ of the underlying bank at time $t$. Under the risk-neutral measure, the implied variance is defined as the price of the contract:
\begin{equation}
\mathbb{IV}\text{ar}_t \equiv e^{-r_t^f} \mathbb{E}_t^Q \left[R_{t+1}^2\right]
\end{equation}
with risk-free rate $r_t^f$. The implied variance, $\mathbb{IV}\text{ar}_t$, measures expected fluctuations in the underlying asset's options contract over a fixed horizon of 30 days. Naturally, this tracks investors' fears that are directly connected to uncertainty regarding next period's expected price movements. Furthermore, \cite{Bakshi2003} suggest that one can use out-of-money (OTM) call and put option prices to compute the implied variance as
\begin{equation}
\label{eq:ivdecomp}
\mathbb{IV}\text{ar}_t = \underbrace{\int_{P_t}^{\infty} \frac{2(1-\log(K/P_t))}{K^2}C(t,t+1,K) dK}_{\mathbb{IV}\text{ar}^{+}_t}  + \underbrace{\int_{0}^{P_t} \frac{2(1+\log(P_t/K))}{K^2}P(t,t+1,K) d K}_{\mathbb{IV}\text{ar}^{-}_t},
\end{equation}
where $C(.)$ and $P(.)$ denote the time $t$ prices of call and put contracts, respectively, with time to maturity of one period and a strike price of $K$. Call option prices reflect a good state of the economy for the stock, while the prices of a put option reflect a bad state of the economy for the stock. The two states are most of the time associated with contrasting investors' beliefs and future expectations \cite[e.g.,][]{buraschi2006}, which will create key components of the network that we aim to build in the next sections. While the equity index OTM puts are usually associated with hedging and insurance against equity market drops \citep[][]{Han2008,Bondarenko2014}, the equity index OTM calls are more commonly associated with optimistic beliefs \citep[][]{buraschi2006}.

Corresponding to an intuitive measure of good and bad events in the stock markets characterized by positive and negative returns, the payoff from the volatility contract can be written as in \cite{kilic2018}:
\begin{equation}
\label{eq:ivdecomp2}
\mathbb{IV}\text{ar}_t \equiv \underbrace{e^{-r_t^f} \mathbb{E}_t^Q \left[R_{t+1}^2 \mathbb{I}\{r_{t+1}>0\}\right]}_{\mathbb{IV}\text{ar}^{+}_t}  + \underbrace{e^{-r_t^f} \mathbb{E}_t^Q \left[R_{t+1}^2 \mathbb{I}\{r_{t+1}\leq 0\}\right]}_{\mathbb{IV}\text{ar}^{-}_t}.
\end{equation}
Intuitively, good and bad components of the payoff add to the total, and the prices of its components can be computed in a model-free way from a bundle of option prices upon a discretization of formula (\ref{eq:ivdecomp}). The total implied variance is the weighted sum of the option prices, and its components are identified by claims that have payoffs related to the sign of the realized return. Hence, good implied variance is identified by call options that pay off only in case the realized return is positive, and bad implied variance is then identified by put options that pay off only if a negative return is realized. To obtain the model-free good and bad implied variance estimates $\widehat{\mathbb{IV}\text{ar}}_t^+$ and $\widehat{\mathbb{IV}\text{ar}}_t^-$ as a discretization of equation (\ref{eq:ivdecomp}), we adopt call and put option prices interpolated around the next 30 days, considering all available strikes for each individual stock options, as detailed in section A in the \oa.

The annualized square roots of the quantities computed for individual companies are then labeled as \vixp and \vixm denoting individual, model-free good and bad implied volatility measures of the expected price fluctuations in the underlying asset's call (put) options over the next month, the pay off of which is related to the positive (negative) return. \vixp captures good fear referring to future uncertainty about the pay off from call options being positive over the next month, and \vixm captures bad fear referring to future uncertainty about the pay off from put options being negative over the next month. In addition, we will work with \vix, which aggregates the good and bad information.

The important characteristics of the network fear connectedness defined in the next section is the predictive power of implied volatility from call and put options that is influenced by the composition of market investors, including speculators, as well as by the demand pressures motivated by hedging or speculation trading activities. \cite{Lemmon2014} found evidence of differences in trading patterns, with individual stock options being mainly driven by unsophisticated investors looking to speculate and index options dominated by hedging demands from sophisticated investors. Speculators may prefer using options for trading to obtain higher leverage \citep{GaoLin}, and they can contribute to reinforcing a financial crisis.
In addition, demand behavior plays important role in explaining the observed implied volatilities \citep{Bollen2004,Garleanu2009}. When option investors receive positive news, the volume of buyer-initiated call trades and/or seller-initiated put trades increases, boosting the implied volatility of call options relative to put options. When option investors receive negative news, the volume of seller initiated call trades and/or buyer initiated put trades increases, which will inflate now the implied volatilities from put options.
Thus, the robustness of the network connectedness measure introduced in this paper depends directly on the robustness of the options market. With the advancement of stock option markets globally, we can envisage that our measures will become valuable tools for central banks and policy makers worldwide.

\subsection{Data on U.S. Financial Institutions} \label{data}

We estimate the good and bad fears of investors about the major financial institutions representing the financial network of the U.S. economy, namely, J.P. Morgan (JPM), Bank of America (BAC), Wells Fargo (WFC), Citigroup (C), Goldman Sachs (GS), Morgan Stanley (MS), U.S. Bancorp (USB), American Express (AXP), PNC Group (PNC) and Bank of New York Mellon (BK). Daily option prices were collected from OptionMetrics\footnote{Data on U.S. stock options are specifically collected from IvyDBUS/v3.1/History/IVYOPPRCD and IvyDBUS/v3.1.1/History/IVYOPPRCD at \url{ftp.ivydb.com}.}, while financial information and market prices are collected from Bloomberg. The dataset ranges from 03-01-2000 to 29-12-2017, covering the 2008 crisis and the remarkable boom that occurred after the crisis. Our sample contains 4528 daily observations for each series. Table A1 in the \oa describes the characteristics of the U.S. financial companies in our sample.

\begin{figure}[ht!]
  \begin{center}
\caption{\textbf{Good and Bad Implied Volatilities of Citigroup}} \label{IVs_Series}
\includegraphics[width=\textwidth]{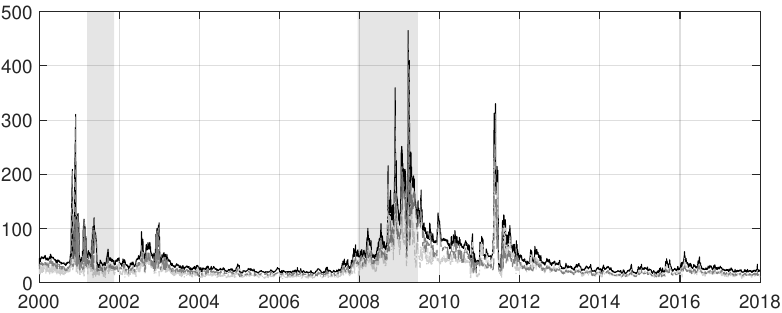}
\caption*{\scriptsize \textit{Notes}: The figure shows the aggregate \vix (black bold), good \vixp (gray dashed) and bad \vixm (black dashed) for Citigroup. The NBER recession periods are highlighted in gray. The selected period spans from 03-01-2000 to 29-12-2017 at a daily frequency.}
\end{center}
\end{figure}

The time dynamics of implied volatility series is illustrated in Figure \ref{IVs_Series} for the Citigroup case. The volatility spikes in alignment with the global financial crisis causing increased uncertainty. High values can also be associated with Citigroup's acquisition of European American Bank and Banamex in mid-2001 and March 2012, when the Federal Reserve reported that Citigroup was one of the few main banks that failed the stress tests. High values mean that investors show the highest levels of fears about future fluctuations in prices connected to corresponding outcome.
Table A2 in the \oa further reports the descriptive statistics for the \vix, \vixp, and \vixm of the ten main financial institutions. We identify Bank of America to show the highest \vix and \vixp average values, followed by Citigroup, while Morgan Stanley presents the highest \vixm average value. On the opposite side, the volatility implied by option prices is lowest for American Express, reaching the lowest minimum values.

The measured volatilities are strongly serially correlated, distributed asymmetrically with strong positive skewness and excess kurtosis, and possibly nonstationary. An approximate normality convenient for further analysis is obtained by taking natural logarithms, and we keep in mind the dependence when building an approximating model. While creating the network of fears in the next section, we will assume that the dynamics come from shifts in the unconditional variances creating nonstationarity. Similarly to \citep{stuaricua2005nonstationarities}, this leads to a convenient approximation of nonstationary data locally by stationary models.

The main drivers of the implied volatilities creating uncertainty in the markets are idiosyncratic news such as merger and acquisitions deals, restructuring and other negotiations. The expectations of investors also react strongly to macroeconomic events such as the dot-com bubble burst, the Enron scandal, the 9/11 terrorist attack, the global financial crisis and the European sovereign debt crisis. A mixture of idiosyncratic and systemic events is found to affect the financial stock implied volatility through the options market, and these shocks then create the network that we aim to measure in the next section.

\section{Asymmetric Network Connectedness of Fears}
\label{meth}

Institutions are connected directly through counterparty risk, contractual obligations or other general business relationships. High-frequency analysis of such networks requires a high-frequency balance sheet and other information, which is generally unavailable. In contrast, option prices and volatility measured in high frequencies reflect the decisions of many agents assessing risks from the existing linkages. The pure market-based approach we use in contrast to other network techniques allows us to monitor the network on a daily frequency as well as to use its forward-looking strength at the cost of minimal assumption.

Working with forward-looking information, we are naturally interested in knowing how a shock to the expected volatility of a stock $j$ will transmit to future expectations about the volatility of a stock $k$. These will define weighted and directed networks. Aggregating this information can provide a systemwide measure of forward-looking connectedness that will measure how strongly investors' expectations are interconnected.

Network connectedness working with causal linkages can be characterized well through variance decompositions from a vector autoregression approximation model \citep{diebold2009, Diebold2012}. Variance decompositions provide useful information about how much of the future variance of variable $j$ is due to shocks in variable $k$. Aggregating variance decompositions yields a simple way to measure how the system is interconnected. \cite{diebold2014} argued that variance decompositions are intimately linked to modern network theory and recently proposed measures of various types of systemic risk, such as marginal expected shortfall \citep{acharya2017} and Delta CoVaR \citep{adrian2016}. Our analysis is also adjacent to that of \cite{song2018}, who developed technical conditions for a network to explain microfinancing decision. Previous literature examined how shocks to volatility measured ex post create linkages in the network. Employing implied volatility measures, we derive informatively different measures of interconnectedness.\footnote{Since our connectedness measures are directly related to key measures of connectedness used in the network literature, hence to systemic risk measures, this study also contributes to the systemic risk literature. Fundamental information transmission from one bank to another has also been considered as a source of banks' connectedness.
Systemic risk may also come from the interaction between asset commonality and funding maturity through an informational channel. This systemic risk is higher, especially when bad information about banks' future solvency arrives in the economy and the asset structures are clustered \citep[see][]{allen2012}. All these market situations can be better understood in a more general framework for banks' information contagion based on volatility since good or bad news in relation to banks influences banks' stock volatility.}

To construct the asymmetric fear connectedness measures, we use the implied volatility indexes computed for the main financial institutions in combination with connectedness measures based on generalized variance decompositions of a vector autoregressive (VAR) approximation model due to \cite{Diebold2012}. In particular, we consider a covariance stationary $N$-variate process $\mathbf{IVIX}_t^{*}=(\mathrm{IVIX}_{1,t}^{*},\ldots,\mathrm{IVIX}_{N,t}^{*})'$ at $t=1,\ldots,T$ described by the VAR model of order $p$ as
\begin{equation}
\mathbf{IVIX}_t^{*} = \boldsymbol\Phi_1 \mathbf{IVIX}_{t-1}^{*} + \boldsymbol\Phi_2 \mathbf{IVIX}_{t-2}^{*} + \ldots + \boldsymbol \Phi_p \mathbf{IVIX}_{t-p}^{*} + \boldsymbol \epsilon_t,
\end{equation}
with $ \boldsymbol \Phi_1,\ldots, \boldsymbol \Phi_p$ coefficient matrices, and $\boldsymbol \epsilon_t$ being white noise with a (possibly nondiagonal) covariance matrix $\boldsymbol \Sigma$. In this model, each variable is regressed on its own $p$ lags, as well as the $p$ lags of all of the other variables in the system; hence, the matrices of the coefficients contain complete information about the connections between variables. It is useful to work with $(N \times N)$ matrix lag-polynomial $\boldsymbol \Phi(L)=[ \boldsymbol I_N - \boldsymbol \Phi_1 L - \ldots - \boldsymbol \Phi_p L^p]$ with $\boldsymbol I_N$ identity matrix, as the model can be written concisely as $\boldsymbol \Phi(L) \mathbf{IVIX}_t^{*} = \boldsymbol \epsilon_t$. Assuming that the roots of $| \boldsymbol \Phi(z) |$ lie outside the unit circle, the VAR process has the following vector moving average (i.e., MA($\infty$)) representation: $\mathbf{IVIX}_t^{*} = \boldsymbol \Psi(L) \boldsymbol \epsilon_t$, where $\boldsymbol \Psi(L)$ matrix of infinite lag polynomials can be calculated recursively from $\boldsymbol \Phi(L) = [\boldsymbol \Psi(L)]^{-1}$ and is key to understanding its dynamics. Since $\boldsymbol \Psi(L)$ contains an infinite number of lags, it must be approximated with the moving average coefficients $\boldsymbol \Psi_h$ calculated at $h=1,\ldots,H$ horizons. The connectedness measures rely on variance decompositions, which are transformations of the $\boldsymbol \Psi_h$ and allow for the measurement of the contribution of shocks to the system.

To construct connectedness measures of aggregate and decomposed good and bad fears, we consider different vectors $\mathbf{IVIX}_t^{*} \in \left\{\mathbf{IVIX}_t,\mathbf{IVIX}_t^{+},\mathbf{IVIX}_t^{-}\right\}$. Since a shock to a variable in the model does not necessarily appear alone, \emph{i.e.}, orthogonally to shocks to other variables, an identification scheme is a crucial step in the calculation of variance decompositions. Standard approaches relying on Cholesky factorization depend on the ordering of the variables and complicate the measures. The generalized identification proposed by \cite{pesaran1998} produces variance decompositions that are invariant to ordering and can be written in the form\footnote{ $(\boldsymbol A)_{j,k}$ denotes the $j$th row and $k$th column of matrix $\boldsymbol A$ denoted in bold. $(\boldsymbol A)_{j,\cdot}$ denotes the full $j$th row; this is similar for the columns. $\sum A$, where $A$ is a matrix that denotes the sum of all elements of the matrix $A$.}
\begin{equation}
	 \boldsymbol \theta^H_{j,k} = \frac{\sigma_{kk}^{-1} \sum_{h = 0}^{H}\left((\boldsymbol \Psi_h \boldsymbol \Sigma)_{j,k}\right)^2}{\sum_{h=0}^{H} (\boldsymbol \Psi_h \boldsymbol \Sigma \boldsymbol \Psi_h')_{j,j}},
	\label{eq:connectedness}
\end{equation} where $\boldsymbol \Psi_h$ is a $(N \times N)$ matrix of moving average coefficients at lag $h$ defined above, and $\sigma_{kk} = \boldsymbol \Sigma_{k,k}$. The $ \boldsymbol \theta^H_{j,k}$ denotes the contribution of the $k$th variable to the variance of forecast error of the element $j$ at horizon $h$. As the rows of the variance decomposition matrix $\boldsymbol \theta^H$ do not necessarily sum to one, each entry is normalized by the row sum as
 $\widetilde{\boldsymbol \theta}^{H}_{j,k} = \boldsymbol \theta^{H}_{j,k}/\sum_{k=1}^N \boldsymbol \theta^{H}_{j,k}.$
Now, the $\sum_{j=1}^N \widetilde{\boldsymbol \theta}^{H}_{j,k}=1$ for any $k$ and the sum of all elements in $\widetilde{\boldsymbol \theta}^{H}$ is equal to $N$ by construction. Note that $\widetilde{\boldsymbol \theta}^{H}_{j,k}$ provides a pairwise measure of connectedness from $j$ to $k$ at horizon $H$. Variance decompositions form a network adjacency matrix defining a weighted, directed network.

The network connectedness measure is then defined as the share of variance in the forecasts contributed by errors other than own errors or as the ratio of the sum of the off-diagonal elements to the sum of the entire matrix \citep{Diebold2012}:
\begin{equation}
  \mathcal{C}^{H} = 100 \cdot \frac{1}{N}\cdot \sum_{1\leq j \neq k\leq N} \widetilde{\boldsymbol \theta}^H_{j,k},
  \label{eq:overallspillover}
\end{equation}
and hence, $\mathcal{C}^{H}$ is the relative contribution to the forecast variances from the other variables in the system.

Similarly to the network aggregate connectedness measure that infers systemwide connectedness, we can define measures that will reveal when an individual bank in the system is a volatility transmitter or receiver. The directional connectedness that measures how much of each bank's $j$ variance is due to other banks $j\ne k$ in the system is given by
\begin{equation}
    \mathcal{C}^H_{j \leftarrow \bullet} = 100 \cdot \frac{1}{N}\cdot \sum_{k=1,j\neq k}^N  \widetilde{\theta}^H_{j,k},
\end{equation}
defining the so-called \textsc{from} connectedness that can be precisely interpreted as from-degrees (often called out-degrees in the network literature) associated with the nodes of the weighted directed network represented by the variance decompositions matrix. Similarly, the contribution of asset $j$ to variances in other variables is computed as
\begin{equation}
    \mathcal{C}^H_{j \to
    \bullet} = 100 \cdot \frac{1}{N}\cdot \sum_{k=1,j\neq k}^N \left( \widetilde{\theta}^H \right)_{k,j},
\end{equation}
and this is the so-called \textsc{to} connectedness. Again, this can be precisely interpreted as to-degrees (often called in-degrees in the network literature) associated with the nodes of the weighted directed network represented by the variance decompositions matrix. These two measures show how other assets contribute to the risk of asset $j$ and how asset $j$ contributes to the riskiness of others. Further, a \textsc{net} connectedness measure showing whether a bank is inducing more risk than it receives from other banks in the system can be calculated as the difference of the directional measures, $\mathcal{C}^H_{j,\textsc{net}} = \mathcal{C}^H_{j \to \bullet} - \mathcal{C}^H_{j \leftarrow \bullet}$. One might also be interested in pairwise relations of risk that can further be described by the \textsc{pairwise} connectedness measure given by $\mathcal{C}^H_{j,k} = 100 \cdot \frac{1}{N}\cdot \left( \widetilde{\theta}^H_{k,j} - \widetilde{\theta}^H_{j,k}\right)$.

To contrast the network connectedness of fundamentally different beliefs revealed by \vixp and \vixm, we define asymmetric fear connectedness. Aggregate fear connectedness ($\mathcal{C}$) and good ($\mathcal{C}^{+}$) and bad ($\mathcal{C}^{-}$) fears in the system can be readily calculated by using appropriate $\mathbf{IVIX}$, $\mathbf{IVIX}^{+}$ and $\mathbf{IVIX}^{-}$ measures. When $\mathcal{C}^{+} \ne \mathcal{C}^{-}$, we have asymmetry in connectedness due to different investors' expectations, which we define as the measure of asymmetric fear connectedness ($\mathcal{AFC}$):
\begin{equation}  \label{SAM}
\mathcal{AFC} = \mathcal{C}^{+} - \mathcal{C}^{-}.
\end{equation}
In other words, when $\mathcal{AFC}>0$, connectedness due to \vixp is greater than connectedness due to \vixm, and vice versa. To shed new light on the nature and sign of the transmitted or received volatility for every financial institution in the system, we compute the directional \textsc{net} as the difference between good \textsc{to} and good \textsc{from} as $\mathcal{C}^{+}_{j,\textsc{NET}} = \mathcal{C}^{+}_{j \to \bullet} - \mathcal{C}^{+}_{j \leftarrow \bullet}$ and between bad \textsc{to} and bad \textsc{from} as $\mathcal{C}^{-}_{j,\textsc{NET}} = \mathcal{C}^{-}_{j \to \bullet} - \mathcal{C}^{-}_{j \leftarrow \bullet}$. Finally, we compute the asymmetric directional \textsc{net} as the difference between $\mathcal{C}^{+}_{j,\textsc{NET}}$ and $\mathcal{C}^{-}_{j,\textsc{NET}}$ as $\mathcal{AFC}_{j,\textsc{NET}} = \mathcal{C}^{+}_{j,\textsc{NET}} - \mathcal{C}^{-}_{j,\textsc{NET}}$.

\subsection{Link to the Network Literature}

The connectedness measures introduced above are intimately related to modern network theory. Algebraically, the adjacency matrix capturing information about network linkages carries all information about the network, and any sensible measure must be related to it. For example, a typical metric used by the wide network literature that provides the user with information about the relative importance or influence of nodes and edges is network centrality. The literature is often also interested in the density describing the proportion of direct ties in a network relative to the total number of the ties. The most useful for our purposes are measures based on the node degree and the closely related concept of network diameter that captures the number of links to other nodes. The distribution shape of the node degrees is closely related to network behavior. As for the connectedness of the network, the location of the degree distribution is key, and hence, the mean of the degree distribution emerges as a benchmark measure of overall network connectedness. Closely related to the idea of distance, the diameter of a network measuring the maximum distance between any two nodes is another measure of network connectedness.

The variance decomposition matrix defining network adjacency matrix is then readily used as a network connectedness that is intimately related to network node degrees and mean degree \citep{diebold2014}. Networks defined by variance decompositions are however more sophisticated than classical network structures. In a typical network, the adjacency matrix is filled with zero and one entries, depending on the node being linked or not, respectively. In the above notion, the variance decompositions can be viewed as a weighted link showing the strength of the connection. In addition, the links are directed, meaning that the $j$ to $k$ link is not necessarily the same as the $k$ to $j$ link, and hence, the adjacency matrix is not symmetric, and so weighted, directed versions of network connectedness statistics can be defined readily including degrees, degree distributions, distances and diameters. Thus, the total directional connectedness measures introduced in the previous section are in-degrees and out-degrees (probability distributions of \textsc{from} or {to} degrees across nodes), and the total connectedness measure is simply the mean degree of the network.

Ultimately, the network connectedness measures based on variance decompositions are tightly linked to and built upon the tradition of dynamic predictive modeling under misspecification pioneered by \cite{white1996estimation}. At the same time, this framework shares similarities with the graphical (network) models contributions focusing on causal linkages as pioneered by \cite{white2009settable}. To capture the causal linkages with the strongly dependent data, one needs to think about more sophisticated tools; hence, the approach unified and conceptualized by \cite{diebold2014} seems appropriate. Using the network topology of good and bad fears later for forecasting, we believe that variance decompositions as a sophisticated network connectedness measure will be more useful than traditional measures due to the reasons discussed.

\section{Fear Connectedness in the Financial Network}  \label{aggreg}

We begin the empirical analysis discussing how individual banks contribute to the aggregate fear connectedness of the network, and we document the time dynamics of connectedness.

\subsection{Directed Network Connectedness of Fear: Static Analysis}
\label{AggregStatic}

The static analysis of the weighted, directed network of fear for the ten main U.S. financial institutions is reported in Table \ref{StaticTot}.\footnote{We use a forecast horizon of 12 days and a VAR order equal to four based on the information criteria. For a dynamic version of the measures, we use a 200-day rolling window. We have also examined the static analysis within a range of different VAR lags and forecast horizons, respectively, such as $p\in\{2,3,4,5\}$, and $h\in\{4,6,10,14\}$, together with different rolling window sizes. The results do not change materially and are available from the authors upon request.} The diagonal values quantify the impact of own shocks to expectations, while off-diagonal elements reveal how fear spreads from one bank to other banks in the financial sector. The directional \textsc{from} connectedness measure documenting vulnerability of the banks to receive shocks from others in the network ranges from 43.93\% for Bank of America to 70.08\% for Bank of New York Mellon, which is the highest receiver. The directional \textsc{to} connectedness measuring the strength of banks transmitting shocks in the bottom row of the table ranges from 19.92\% for Bank of America to 108.85\% for Goldman Sachs, the latter being identified as the largest transmitter, which in line with the literature identifying Goldman Sachs as being systematically important using ex post measures \citep[see][]{hautsch2014,geraci2018}.
\begin{table} [ht!]
  \centering
  \caption{\textbf{Static Fear Connectedness in the Financial Sector}}   \label{StaticTot}
  \begin{threeparttable}
  \scriptsize{
  \centering
  \begin{tabular}{lccccccccccc}
  \toprule
  \multicolumn{12}{c}{\vix Connectedness}\\
  \cmidrule{2-11}
   &JPM & BAC& WFC &	CITI & GS	&MS&	USB&	AXP&	PNC&	BK& FROM\\
   \cmidrule{2-11}
JPM & 	\Gray{46.98}&	2.16&	3.58&	3.26&	10.36&	9.46&	8.69&	6.63&	2.06&	6.77&	53.01\\
BAC	&4.03&	\Gray{56.06}&	3.04&	3.28&	8.20&	7.05&	5.67&	4.99&	1.65&	5.98	&43.93\\
WFC	&6.93&	1.78&	\Gray{44.41}&	4.61&	11.11&	6.37&	7.19&	5.63&	5.66&	6.26	&55.58\\
CITI&	5.23&	2.09&	3.44&	\Gray{54.67}&	8.92&	7.20&	5.69&	5.44&	2.33&	4.94	&45.32\\
GS&	6.34&	2.25&	4.81&	5.54&	\Gray{35.55}&	18.29&	9.35&	7.36&	2.29&	8.19	&64.44\\
MS&	6.91&	2.58&	4.12&	5.11&	20.29&	\Gray{34.34}&	8.59&	6.40&	2.86&	8.74	&65.65\\
USB&	8.23&	2.26&	4.97&	4.34&	13.62&	9.67&	\Gray{36.33}&	7.67&	3.48&	9.39	&63.66\\
AXP	&7.14&	2.51&	4.21&	4.45&	12.65&	11.47&	10.42&	\Gray{35.71}&	3.02&	8.37	&64.28\\
PNC	&5.82&	1.34&	4.93&	3.11&	8.69&	7.53&	6.85&	4.22&	\Gray{49.89}&	7.57	&50.10\\
BK	&7.19&	2.90&	4.25&	4.21&	14.97&	14.50&	10.38&	7.77&	3.87&	\Gray{29.91}	&70.08\\
\cmidrule{2-11}
TO&	57.86&	19.92&	37.38&	37.96&	108.85&	91.57&	72.87&	56.14&	27.25&	66.25&	TOTAL\\
NET&	4.84&	-24.01&	-18.19&	-7.35&	44.40&	25.92&	9.21&	-8.13&	-22.84&	-3.82&	\textbf{57.61}\\
  \bottomrule
  \end{tabular}
  \caption*{\scriptsize \textit{Notes}: The table contains a decomposition of forecast error variance computed for the aggregate \vix indexes for the ten main U.S. banks. Elements in the off-diagonal entries are the \textsc{pairwise} directional connectedness, while the diagonal elements (in gray) are the banks' own variance. The off-diagonal row and column sum to \textsc{to} and \textsc{from} connectedness, respectively. The \textsc{net} row at the bottom is the difference between \textsc{to} and \textsc{from}. The bottom-right element is the total connectedness index in the system. The selected time period spans from 03-01-2000 to 29-12-2017.}}
  \end{threeparttable}
\end{table}
The \textsc{pairwise} values in the off-diagonal entries indicate the directional connectedness between the corresponding companies. The highest \textsc{pairwise} connectedness emerges from Goldman Sachs to Morgan Stanley, with 20.29\% of the Morgan Stanley's future variation due to the shocks from Goldman Sachs. The second-highest number is in the opposite direction, with 18.29\% due to the shocks from Morgan Stanley; hence, this pair seems to create the strongest bidirectional connection. The positive values of the \textsc{net} connectedness point to fear transmitters, while the negative values identify fear receivers in the system, with Goldman Sachs (44.4\%) and Morgan Stanley (25.92\%) being the main fear transmitters in the system. The total fear connectedness being 57.61\% documents a rather strongly connected network of fears in the financial system.

\subsection{Directed Network Connectedness of Fear: Time Dynamics}
\label{aggregDynamic}

With the dynamic evolution of the markets, we expect the network to also show strong time dynamics.
\begin{figure}[ht!]
 \begin{center}
\caption{\textbf{Total Fear Connectedness Index}} \label{SI_TOT_VIX}
\includegraphics[width=\textwidth, height=0.3\textheight]{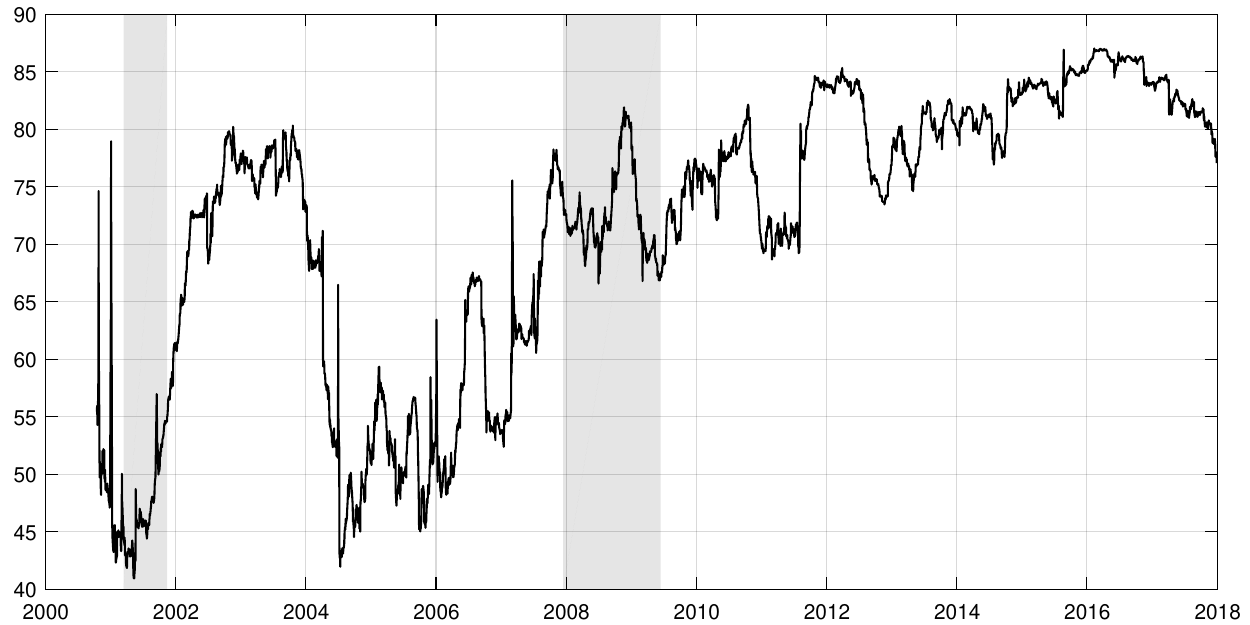}
\caption*{\scriptsize \textit{Notes}: This figure shows the total fear connectedness for the ten main financial institutions' aggregate \vix indexes. The NBER recession periods are highlighted in gray. The selected period spans from 03-01-2000 to 29-12-2017 at a daily frequency. }
\end{center}
\end{figure}
Figure \ref{SI_TOT_VIX} illustrates how the total fear connectedness index spiked twice in the early 2000s due to several specific pieces of news and M\&A deals and to the burst of the dot-com bubble in March 2000. These events, in addition to others, such as the 9/11 terrorist attack and the Enron and MCI WorldCom scandals, are found to have increased the total fear connectedness index at the end of 2001 from 45\% to 75\% in only one year. The index remained at high levels, close to 80\%, for a few years until it decreased in mid-2004. This period was followed by several smaller cycles corresponding to the U.S. tightening its monetary policy and increases in long-term interest rates. The total connectedness index rose again in February 2007 in alignment with the beginning of the subprime crisis. After decreasing for few months, it jumped up in mid-2007, increasing by more than 20\% to levels near 80\%. In the middle of the global financial crisis, the index spiked again in accordance with the losses of Merrill Lynch and the collapse of Lehman Brothers in September 2008, documenting the increasing strength of the network of fears.

It is interesting to note that the connectedness of fears was almost gradually increasing from 2004 with few local peaks corresponding to the Great Financial Crisis and Eurozone sovereign debt crisis (May 2010 and August 2011). We notice that the index level remained high thereafter, pointing to increased uncertainty, even during more tranquil times such as in periods after the global financial crisis. Previous studies using historical volatility measures documented a high level of connectedness after the global financial crisis and European sovereign debt crisis when looking at firms within the U.S. financial sector \cite[see][]{diebold2015,barunik2018}. We notice a peak in the fear connectedness after those events, with the average value being above the precrisis range. This might be due to the fact that we are looking at the network of expectations about future volatility based on ex ante measures in contrast with the previous literature that used ex post measures. The option trading activity (e.g., speculative activity) might also play a role in increasing the forward-looking volatility connectedness within the financial system. Hence, the dynamics of connectedness show that shocks to expectations about future uncertainty play an increasingly important role in the financial network.

\section{Asymmetric Fear Connectedness in the Financial Network}
\label{AsymmetryVol}

\subsection{Asymmetric Fear Connectedness: Static Analysis}
\label{asymmstatic}

Moving to the network of good and bad fears, we first look at the unconditional analysis reported in Table \ref{StaticDecomp}. We document weaker connections in both networks of good and bad fears, meaning that shocks to expectations about uncertainty connected to positive and negative returns create weaker networks. We find that Morgan Stanley is most vulnerable to shocks to both good fears (46.84\%) and bad fears (34.97\%), while Goldman Sachs is found to be the largest transmitter of good fears (61.02\%) and bad fears (55.01\%), and the pair shows the highest good and bad \textsc{pairwise} connectedness.
\begin{table} [ht!]
  \centering
  \caption{\textbf{Asymmetric Fear Connectedness in the Financial Sector}}   \label{StaticDecomp}
  \begin{threeparttable}
  \scriptsize{
  \centering
  \begin{tabular}{lccccccccccc}
  \toprule
  \multicolumn{12}{c}{\vixp Connectedness}\\
   \cmidrule{2-11}
   &JPM & BAC& WFC &	CITI & GS	&MS&	USB&	AXP&	PNC&	BK& FROM\\
   \cmidrule{2-11}
   JPM	&\Gray{73.69}&	0.79&	1.55&	1.67&	4.11&	5.28&	5.53&	2.79&	0.73&	3.82&	 26.30\\
    BAC	&0.90&	\Gray{79.51}&	1.85&	3.36&	2.28&	2.52&	0.80&	3.63&	0.90&	4.20&	20.48\\
    WFC	&3.04&	0.88&	\Gray{72.25}&	3.53&	6.20&	1.39&	3.80&	4.13&	1.01&	3.70&	27.74\\
    CITI&	0.72&	2.05&	2.10&	\Gray{80.67}&	4.04&	1.73&	1.48&	2.79&	1.88&	2.49&	19.32\\
    GS&2.50&	1.09&	4.78&	4.05&	\Gray{58.31}&	12.18&	5.95&	5.31&	1.15&	4.63&	41.68\\
    MS&	4.05&	1.76&	2.39&	3.56&	17.76&	\Gray{53.15}&	4.03&	3.64&	1.88&	7.72&	46.84\\
    USB&	4.72&	0.39&	4.00&	2.94&	7.15&	2.66&	\Gray{63.83}&	5.59&	2.53&	6.13&	36.16\\
    AXP	&2.70&	1.70&	3.90&	2.83&  7.81&	4.79&	5.98&	\Gray{63.34}&	1.95&	4.96&	36.65\\
    PNC&	2.61&	0.92&	2.77&	3.78&	3.92&	3.80&	4.90&	4.05&	\Gray{68.42}&	4.77&	31.57\\
    BK&	3.67&	2.17&	3.46&	2.61&	7.71&	10.08&	6.52&	5.98&	2.90&	\Gray{54.85}&	45.14\\
    \cmidrule{2-11}
    TO&	24.96&	11.81&	26.85&	28.37&	61.02&	44.46&	39.02&	37.95&	14.97&	42.45&	TOTAL\\
    NET&	-1.33&	-8.66&	-0.88&	9.05&	19.33&	-2.37&	2.85&	1.29&	-16.59&	-2.68&	\textbf{33.19}\\
  \midrule
  \multicolumn{12}{c}{\vixm Connectedness}\\
   \cmidrule{2-11}
   &JPM & BAC& WFC &	CITI & GS	&MS&	USB&	AXP&	PNC&	BK& FROM\\
   \cmidrule{2-11}
  JPM&	\Gray{73.08}&	1.50&	0.54&	3.86&	5.04&	3.45&	6.28&	3.97&	0.69&	1.54&	26.91\\
BAC&	0.89&	\Gray{80.11}&	3.92&	2.68&	2.57&	4.19&	2.79&	0.45&	0.91&	1.45&	19.88\\
WFC&	1.79&	2.75&	\Gray{73.53}&	4.57&	5.42&	1.83&	2.08&	2.69&	2.99&   2.29&	26.46\\
CITI&	2.43&	2.33&	2.82&	\Gray{80.73}&	2.42&	2.57&	1.71&	2.10&	1.09&	1.76&	19.26\\
GS&	    1.43&	0.91&	3.40&	2.80&	\Gray{71.83}&	10.39&	4.13&	3.01&	0.58&	1.48&	28.16\\
MS&	    2.85&	2.24&	2.25&	3.71&	15.69&	\Gray{65.02}&	2.90&	1.32&	0.83&	3.15&	34.97\\
USB&	6.54&	1.89&	1.69&	3.07&	6.43&	3.33&	\Gray{71.05}&	4.05&	0.68&	1.20&	28.94\\
AXP&    3.13&	0.24&	3.20&	4.96&	8.58&	2.30&	6.58&	\Gray{67.32}&	1.80&	1.84&	32.67\\
PNC&	2.57&	1.48&	2.75&	1.61&	4.77&	0.94&	1.32&	2.49&	\Gray{80.72}&	1.28&	19.27\\
BK &    1.23&	2.60&	3.79&	3.61&	4.04&	5.79&	1.47&	1.82&	1.27&	\Gray{74.32}& 25.67 \\
\cmidrule{2-11}
TO&	22.92&	15.99&	24.37&	30.92&	55.01&	34.84&	29.29&	21.94&	10.87&	16.03&	TOTAL\\
NET&	-3.98&	-3.89&	-2.08&	11.65&	26.84&	-0.12&	0.35&	-10.72&	-8.39&	-9.64&	\textbf{26.22}\\
\bottomrule
  \end{tabular}
    \caption*{\scriptsize \textit{Notes}: The table contains the adjacency matrix of the good and bad fear networks as forecast error variance decompositions computed for the \vixp and\vixm for the ten main U.S. banks. The elements in the off-diagonal entries are the \textsc{pairwise} directional connectedness, while the diagonal elements (in gray) are banks' own variance. The off-diagonal row and column sum to \textsc{to} and \textsc{from} directional connectedness, respectively. The \textsc{net} row at the bottom is the difference between \textsc{to} and \textsc{from}. The bottom-right element is the total connectedness index in the system. The selected time period spans from 03-01-2000 to 29-12-2017.}}
  \end{threeparttable}
\end{table}
Since Goldman Sachs plays a pivotal role in the financial sector, being an aggregate transmitter regardless of the nature of the uncertainty, we also document interesting dynamics for other banks switching roles from \textsc{net} receiver to \textsc{net} transmitter, or vice versa, confirming asymmetries in the transmission mechanism. PNC Bank, Bank of America and Bank of New York Mellon are found to be the weakest banks, given that they receive volatility from the system regardless of the volatility measure.

\subsection{Asymmetric Fear Connectedness: Time Dynamics}
\label{AsymmDynamic}

\begin{figure}[ht!]
 \begin{center}
\caption{\textbf{Fear Connectedness Indexes and AFC}} \label{IV_SI&SAM}
\includegraphics[width=\textwidth]{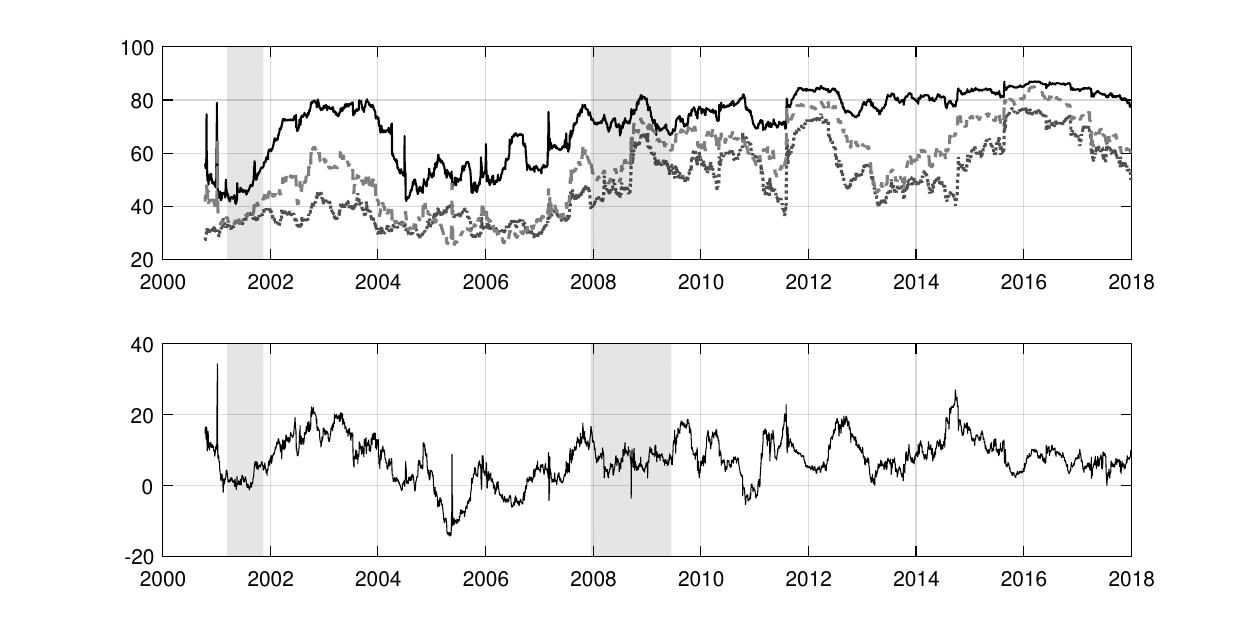}
\caption*{\scriptsize \textit{Notes}: The upper plot shows the comparison between aggregate ($\mathcal{C}$), good ($\mathcal{C}^{+}$) and bad ($\mathcal{C}^{-}$) connectedness measures plotted by a black solid line, gray dashed line and black dotted line, respectively. The bottom plot shows the asymmetric fear connectedness (AFC). The NBER recession periods are highlighted in gray. The selected period spans from 03-01-2000 to 29-12-2017 at a daily frequency.}
\end{center}
\end{figure}

The evolution of good and bad fear network connectedness and asymmetric fear connectedness is depicted in Figure \ref{IV_SI&SAM}, confirming that in the financial sector, good fears are more strongly connected than bad fears for the entire period. However, in some specific periods, such as during the two recessions and during the Eurozone sovereign debt crisis, bad fear connectedness also increased.
Connectedness due to shocks in good fears spike before the 2000-2001 dot-com bubble, dragging up the aggregate connectedness, and dropping quickly after the bubble burst. Shocks to bad fears create less strong connections in comparison to shocks to good fears before a recession (almost 50\% difference in 2002-2003), while they play an equal role during the recession.

With a rather high correlation of the good and bad connectedness measures of 90\%, it is interesting to note that both series strongly co-move and tend to deviate from the aggregate connectedness in some periods. The peak in bad connectedness during 2005 can be attributed to extreme uncertainty about a possible U.S. housing bubble burst, reflected in the stronger connections of expectations of put option buyers. More interestingly, both measures increased well before the crisis during 2007, when their values almost doubled. A similar situation occurred at the beginning of 2011 in accordance with the Eurozone sovereign debt crisis, while the last increase in 2014-2015 is followed by the decreasing strength of the connections.

The connectedness of bad fears is stronger than that of good fears only during 2005, 2006 and 2011. In these periods, shocks to the expectations of put option buyers create a stronger network. Put options reflect investors' expectations about uncertainty connected to decreasing prices and future financial and economic downturns since they are traded as insurance assets \citep[e.g.,][]{Bollen2004,Ang2006,Bondarenko2014}. Hence, the connectedness of bad fears ($\mathcal{C}^{-}$) appears to be a useful monitoring tool candidate, and we aim to explore it in further sections.

\subsection{Case Study: Goldman Sachs}
\label{zoomin}

\begin{figure}[ht!]
 \begin{center}
\caption{\textbf{Good and Bad Net Fear Connectedness -- Goldman Sachs}} \label{GS_Net&SAM}
\includegraphics[width=\textwidth, height=0.3\textheight]{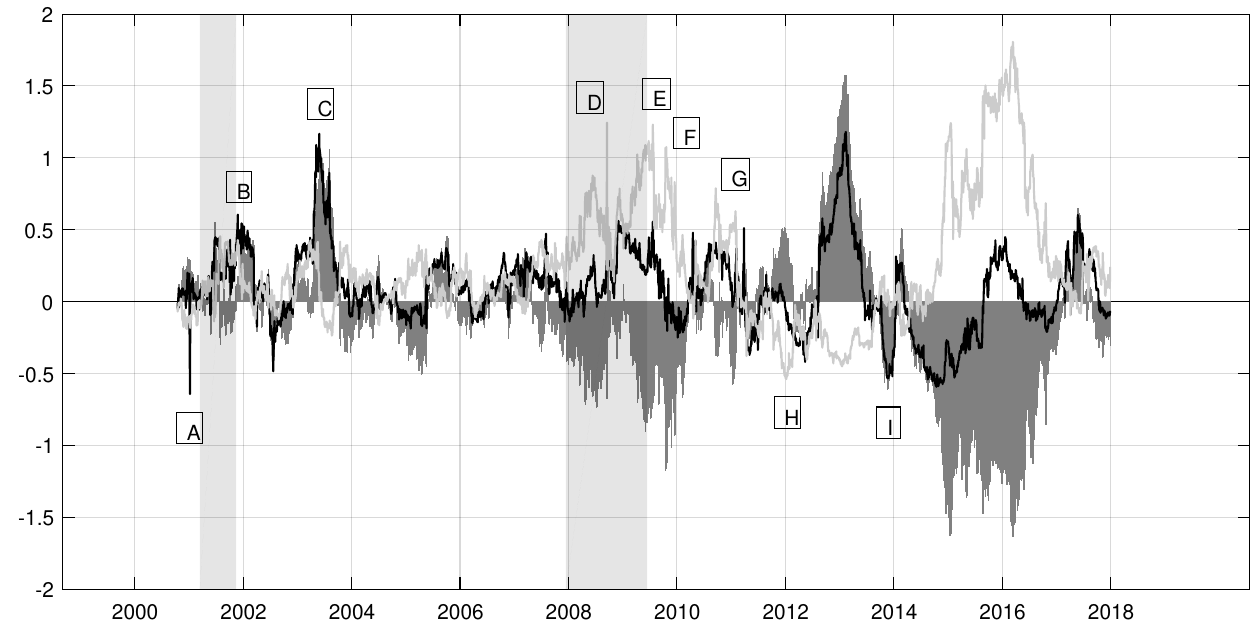}
\caption*{\scriptsize \textit{Notes}: The figure shows the \textsc{net} good fear connectedness, $\mathcal{C}^+_{GS,\textsc{net}}$, and \textsc{net} bad fear connectedness, $\mathcal{C}^-_{GS,\textsc{net}}$ for Goldman Sachs, together with the AFC, computed as the difference between the two. The main specific company events during the time period are reported as follows: [A] First Internet Bond Offering [B] Spear, Leeds and Kellogg Acquisitions [C] Joint Venture with JBWere [D] Short-Selling of Subprime Mortgage-Backed Securities [E] \$10 Bn. Preferred Stock from TARP [F] TARP Repayment [G] JBWere Full Control [H] Global Alpha Shutdown [I] \$17 Bn. Bond Offering by Apple Inc. The NBER recession periods are highlighted in gray. The selected period spans from 03-01-2000 to 29-12-2017 at a daily frequency.}
\end{center}
\end{figure}

In addition to the time evolution of connectedness, it is also interesting to study the role of institutions in the network. We discuss the dynamics of a representative bank, Goldman Sachs, which is the main transmitter of fears in the system.\footnote{A detailed ranking of the institutions as \textsc{net} good or \textsc{net} bad fear transmitters or receivers is presented in section B of the \oa. The \oa also contains similar case studies of other financial institutions found to be the top \textsc{net} aggregate fear transmitters and receivers in our previous analysis (section C). The same notation will apply for these banks, and major specific company events, along with systematic events, will be reported for the selected time period.}  Figure \ref{GS_Net&SAM} illustrates the measures denoting the \textsc{net} good fear as $\mathcal{C}^+_{GS,\textsc{NET}}$ and the \textsc{NET} bad fear as $\mathcal{C}^-_{GS,\textsc{NET}}$.

In January 2000, Goldman Sachs and Lehman Brothers were the lead managers of the first internet bond offering for the World Bank, which is found to correspond to one of the highest levels of $\mathcal{C}^+_{GS,\textsc{NET}}$ received by Goldman Sachs. When Goldman Sachs purchased Spear, Leeds and Kellogg in September 2000 for more than \$6 billion, significant good and bad fear was transmitted into the system. In 2003, Goldman Sachs took an almost 50\% stake in a joint venture, together with JBWere, which resulted in a spike of $\mathcal{C}^+_{GS,\textsc{NET}}$ transmitted, together with an increase in absorbed $\mathcal{C}^-_{GS,\textsc{NET}}$. We find $\mathcal{C}^-_{GS,\textsc{NET}}$ transmission during the financial crisis, especially in 2007, when Goldman Sachs' traders bet against the mortgage market, which exhibited an alarming pessimistic signal to the U.S. financial sector.
In October 2008, Goldman Sachs received a \$10-billion preferred stock investment from the U.S. Treasury as part of the Troubled Asset Relief Program (TARP). This bailout intervention appeared to increase the instability of the U.S. financial sector, resulting in an increase in transmitted $\mathcal{C}^-_{GS,\textsc{NET}}$.
In June 2009, Goldman Sachs repaid the U.S. TARP investment, resulting in a drop in the transmitted \textsc{net} $\mathcal{C}^-_{GS,\textsc{NET}}$ but also in an increase in the received $\mathcal{C}^+_{GS,\textsc{NET}}$ as a sign of recovery. One of the highest peaks of $\mathcal{C}^+_{GS,\textsc{NET}}$ received is found in April 2013, when Goldman Sachs, together with Deutsche Bank, led a \$17 billion bond offering by Apple Inc. During the same year, Goldman Sachs led Twitter's IPO. Both IPOs resulted in a stable $\mathcal{C}^+_{GS,\textsc{NET}}$ reception and a $\mathcal{C}^-_{GS,\textsc{NET}}$ transmission from that time onward.

Note that $\mathcal{C}^+_{GS,\textsc{NET}}$ measures are computed as a difference between good \textsc{to} and \textsc{from} (transmitters and receivers of good fear), implying that \textsc{net} measures reflect not only systematic events related to the bank but also possibly other events with respect to others in the system. For this reason we also study the predictive power of the measures with respect to banks' performance and show the usefulness of the decomposed \textsc{net} measures from a microeconomic point of view.\footnote{Due to page constraints, we relegate this analysis to section D of the \oa.} The next section focuses on the predictive ability of the network connectedness measures from a macroeconomic point of view.

\section{Predictive Power of Fear Connectedness Measures}
\label{predictive}

Our main interest in building the ex ante network connectedness measures is to see if they are helpful in predicting future macroeconomic conditions, as well as the potential increase in economic uncertainty. Our hypothesis is that forward-looking connectedness measures may result in an \textit{early} warning tool to forecast declines in the U.S. macroeconomic conditions or increases in financial and economic uncertainty. We select the following monthly indicators, which reflect the macroeconomic and economic conditions, such as the Aruoba-Diebold-Scotti (ADS) Business Condition Index \citep{aruoba2009}, the Chicago FED National Activity Index, CFNAI, the Kansas City Financial Stress Index (KCFSI) \citep[see][]{hakkio2009}, the NBER recession period dummy variable and the U.S. Industrial Production (IP). As uncertainty proxies, we select the Economic Policy Uncertainty (EPU) index \citep[see][]{Baker2016}, the GeoPolitical Risk (GPR) index by \cite{caldara2018}, the Economic Uncertainty Index (EUI) by \cite{bali2014}, the Chicago Board of Exchange (CBOE) VIX, and the average conditional volatility based on GARCH(1,1) of some U.S. macroeconomic variables (AVGVOL). They are collected according to their available sample at a monthly frequency.\footnote{A more detailed description of the selected variables and their sources is provided in section E1 in the \oa.}

The connectedness measures for this exercise are computed every month by aggregating three months of implied volatility observations (60 trading days) with regard to the indexes, \vix, \vixm and \vixp, for every financial institution.\footnote{For the main analysis, we compute the connectedness measures with a number of VAR lags equal to 1. Different VAR lags choice does not materially change our results. The results are available from the authors upon request.} The quarterly connectedness measures are rolled every month to produce monthly $\mathcal{C}_t$, $\mathcal{C}_t^{+}$ and $\mathcal{C}_t^{-}$ observations that reflect the previous quarter. The monthly macroeconomic and uncertainty indicators are taken as the average of the previous quarter, and recession is marked binary as 1 when the average obtains values $>0.5$ and 0 when the average obtains values $<0.5$. This process allows us to match the information from the fear connectedness indexes with the macro and uncertainty indicators, thus creating monthly observations that reflect the information in the previous quarter.

Our main hypothesis is that the forward-looking measures of the network are informative for the future of both economic conditions and economic uncertainty. Further, the decomposed connectedness measures $\mathcal{C}_t^{-}$ and $\mathcal{C}_t^{+}$ may carry additional information compared to the aggregate fear index $\mathcal{C}_t$. To test these hypotheses, we investigate both the in-sample and out-of-sample predictive power of the fear networks with the main focus being on the latter.


We run the following predictive equations:
\begin{equation} \label{Predictaggregate}
\mathrm{X}_{t+h} = \beta_0 + \beta \mathcal{C}_t + \sum_{k=0}^{11} \gamma_k \mathrm{X}_{t-k} + \epsilon_t
\end{equation}
\begin{equation}  \label{Predictdecomp}
\mathrm{X}_{t+h} = \beta_0  + \beta^- \mathcal{C}_t^{-} + \beta^+ \mathcal{C}_t^{+} + \sum_{k=0}^{11} \gamma_k \mathrm{X}_{t-k} + \epsilon_t, \end{equation}
where $\mathrm{X} \in \{\mathrm{ADS, CFNAI, KCSFI, NBER,IP,EPU, GPR, EUI, VIX, AVGVOL}\}$ is one of the  macroeconomic and economic uncertainty indicators for $h=1,...,12$-step-ahead horizons up to one year. The traditional aggregate variables used as indicators are highly persistent with high first-order autocorrelations, alleviating concerns about the estimates. Hence, we add the lags of the indicators to control for persistence.
We are mainly interested in finding if the networks are informative for the $h=1,...,12$-step predictions; we include up to 12 lags of the dependent variables in the predictive regressions and show whether or not our connectedness measures are still found to be significant, even after taking into account these control variables. A similar approach has been taken by \cite{allenbali2012} and \cite{almeida2017}; hence, the choice of the regressions makes our analysis directly comparable to the literature.
For the NBER recession dichotomous variable, a probit regression is fitted, while for the other variables, least squares regressions are estimated. Bootstrapped standard errors are reported in the tables with respect to the in-sample analysis.\footnote{We are aware of the possible issue concerning ``generated'' regressors in our in-sample analysis, requiring adjustments to the standard errors. The network measure dimension $N$ is relatively small, and the regularity conditions of \cite{bai2006} may not hold since the least squares estimates are $\sqrt{T}$ consistent and asymptotically normal if $\sqrt{T/N} \rightarrow 0$. Therefore, we estimate our standard errors in the in-sample analysis using a bootstrapping approach \citep[see][]{gospodinov2013}.}

The impact of the fear connectedness of the financial sector on macroeconomic conditions and uncertainty measures in terms of $\beta$, $\beta^-$ and $\beta^+$ coefficients is summarized in section E2 in the \oa for the sake of space. With respect to the in-sample analysis, our findings indicate that even after controlling for 12 lags of the endogenous variables, the $\beta^-$ and $\beta^+$ coefficients of the decomposed connectedness measures are found to be significantly different from zero in a number of cases, showing greater strength in comparison to the aggregate connectedness captured by $\beta$.
We observe that the connectedness of both bad ($\mathcal{C}_t^{-}$) and good ($\mathcal{C}_t^{+}$) fear in financial sector has additional explanatory power for ADS Business Condition Index, CFNAI Index of National Activity and U.S. Industrial Production index approximately a quarter to a year in advance. In addition, the connectedness of bad fears can signal a recession, providing an early-warning alarm a few months before the aggregate $\mathcal{C}_t$, emphasizing the importance of investors' expectations contained in the put options. We also perform the same predictive exercise by grouping the information contained in $\mathcal{C}_t^{-}$ and $\mathcal{C}_t^{+}$ as a ratio between the two. We still achieve similar findings with a more compact and parsimonious equation. We report the set of in-sample predictability results in section E2 and E3 in the \oa. In the next subsection, we focus on the usefulness of our networks for out-of-sample predictability.

\subsection{Out-of-Sample Prediction of Macroeconomic Conditions and Uncertainty}

A natural way to assess the out-of-sample predictability power of our networks is to compare the full models, including the connectedness measures with restricted models, where $\beta=\beta^-=\beta^+=0$. In case the network measures are informative about predictions, we should document significant improvement of the out-of-sample errors from the full model in comparison to the restricted model accompanied by the nonzero in-sample estimates of these coefficients.

Keeping the precrisis period of 2000-2007 as the in-sample period, we estimate the coefficients of the models and use the rest of the sample for out-of-sample comparison of the forecast errors. The choice of the predictive regressions is mainly driven by parsimony.
We compare their out-of-sample performance with a pure autoregressive structure, with the aim of checking whether or not the addition of our network measures increase predictability.\footnote{As a robustness check, we have also checked the performance of our models compared to a simple historical average (rolled every quarter as to match the information content of our network measures). The results are reported in section E4 of the \oa.}

Since we are comparing nested models, we use the mean square forecast error (MSFE)-adjusted statistics of \cite{clark2007} for comparing the nested model forecasts, which performs well in finite samples. The (MSFE)-adjusted statistics test the null hypothesis that the restricted model average MSFE is less than or equal to the full model MSFE against the one-sided (upper-tail) alternative hypothesis that the benchmark restricted model error is greater than the error from full model. This corresponds to the null hypothesis, that information contained in the network does not improve forecasts in terms of errors.

\begin{table}[hbt!]
 \centering
 \caption{\textbf{Out-of-Sample Macroeconomic and Uncertainty Indicator Prediction}} \label{OOSPredict1}
 \begin{threeparttable}
 \centering
 \scriptsize{
 \begin{tabular}{crlccccccccc}
   \toprule
     & & &\multicolumn{1}{c}{ADS} && \multicolumn{1}{c}{CFNAI} &&  \multicolumn{1}{c}{KCSFI} &&   \multicolumn{1}{c}{NBER} &&  \multicolumn{1}{c}{IP}\\
 \cmidrule{1-2} \cmidrule{4-4} \cmidrule{6-6} \cmidrule{8-8} \cmidrule{10-10} \cmidrule{12-12}
 Horizon & Predictor && 	MSFE adj. & &  MSFE adj. & &  	MSFE adj. & &		MSFE adj.	& & MSFE adj.	\\
 \cmidrule{1-2} \cmidrule{4-4} \cmidrule{6-6} \cmidrule{8-8} \cmidrule{10-10} \cmidrule{12-12}

1 &  $\mathcal{C}_t$ &&	0.69&&	1.90*&&	0.88&&	1.25&&		0.49\\
 &  $\mathcal{C}_t^{-}$ \& $\mathcal{C}_t^{+}$ &&	0.50&&	2.05**&&	0.60&&	2.45***&&	1.30*\\
 \cmidrule{1-2} \cmidrule{4-4} \cmidrule{6-6} \cmidrule{8-8} \cmidrule{10-10} \cmidrule{12-12}

3 &  $\mathcal{C}_t$ &&	0.89&&	1.93**&&	1.20&&	1.27*&&	0.20\\
 &  $\mathcal{C}_t^{-}$ \& $\mathcal{C}_t^{+}$ &&	1.09&&	2.27***&&	0.55&&	1.59**&&	1.26*\\
 \cmidrule{1-2} \cmidrule{4-4} \cmidrule{6-6} \cmidrule{8-8} \cmidrule{10-10} \cmidrule{12-12}

6 &  $\mathcal{C}_t$ &&	0.49&&	2.03**&&	0.84&&	1.31&&	0.22\\
 &  $\mathcal{C}_t^{-}$ \& $\mathcal{C}_t^{+}$ &&	0.50&&	2.32***&&	0.87&&	-0.58&&	0.95\\
 \cmidrule{1-2} \cmidrule{4-4} \cmidrule{6-6} \cmidrule{8-8} \cmidrule{10-10} \cmidrule{12-12}

12 &  $\mathcal{C}_t$ &&	0.56&&	1.81*&&	0.52&&	1.52*&&	0.38\\
 &  $\mathcal{C}_t^{-}$ \& $\mathcal{C}_t^{+}$ &&	0.38&&	2.14**&&	0.30&&	1.26*&&	2.03**\\
\toprule
    & & &\multicolumn{1}{c}{EPU} && \multicolumn{1}{c}{GPR} & & \multicolumn{1}{c}{EUI} & &  \multicolumn{1}{c}{VIX} & & \multicolumn{1}{c}{AVGVOL}\\
     \cmidrule{1-2} \cmidrule{4-4} \cmidrule{6-6} \cmidrule{8-8} \cmidrule{10-10} \cmidrule{12-12}
 Forecast & Predictor && 	MSFE adj. & &  MSFE adj. & &  	MSFE adj. & &		MSFE adj.	& & MSFE adj.	\\
     \cmidrule{1-2} \cmidrule{4-4} \cmidrule{6-6} \cmidrule{8-8} \cmidrule{10-10} \cmidrule{12-12}
1 &  $\mathcal{C}_t$ &&	2.78***&&	3.29***&&	1.64*&&	3.03***&&	-1.32\\
 &  $\mathcal{C}_t^{-}$ \& $\mathcal{C}_t^{+}$ &&	2.64***&&	2.82***&&	1.35*&&	3.09***&&	-1.28\\
     \cmidrule{1-2} \cmidrule{4-4} \cmidrule{6-6} \cmidrule{8-8} \cmidrule{10-10} \cmidrule{12-12}

3 &  $\mathcal{C}_t$ &&	3.15***&&	3.27***&&	1.63*&&	3.21***&&	-1.31\\
 &  $\mathcal{C}_t^{-}$ \& $\mathcal{C}_t^{+}$ &&	3.06***&&	2.98***&&	1.71**&&	2.69***&&	-0.78\\
     \cmidrule{1-2} \cmidrule{4-4} \cmidrule{6-6} \cmidrule{8-8} \cmidrule{10-10} \cmidrule{12-12}

6 &  $\mathcal{C}_t$ &&	3.03***&&	3.25***&&	1.52*&&	3.08***&&	-1.37\\
 &  $\mathcal{C}_t^{-}$ \& $\mathcal{C}_t^{+}$ &&	2.52***&&	2.77***&&	1.65**&&	2.77***&&	-1.09\\
     \cmidrule{1-2} \cmidrule{4-4} \cmidrule{6-6} \cmidrule{8-8} \cmidrule{10-10} \cmidrule{12-12}

12 &  $\mathcal{C}_t$ &&	2.65**&&	3.09***&&	1.61*&&	2.86***&&	-1.35\\
 &  $\mathcal{C}_t^{-}$ \& $\mathcal{C}_t^{+}$ &&	2.09**&&	2.94***&&	1.87**&&	2.34***&&	-1.46\\
\bottomrule
\end{tabular}
\begin{tablenotes}
\item {\scriptsize \textit{Notes}: The table presents the \cite{clark2007} mean square forecast error (MSFE)-adjusted statistic comparing the out-of-sample predictions from the full models with our connectedness measures and restricted model with $\beta=\beta^-=\beta^+=0$. The in-sample period is between 2000-2007, while the rest is considered as the out-of-sample evaluation forecast period. The results are reported for the forecast horizons $\in \{1,3,6,12\}$. Rejections of the null hypothesis, that the full model containing network information $\mathcal{C}_t$ or $\mathcal{C}_t^{-}$ and $\mathcal{C}_t^{+}$ does not improve the predictions, are reported as ${*}$, ${**}$, and ${***}$, for the $10\%$, $5\%$, and $1\%$ significance levels, respectively.}
\end{tablenotes}
}
\end{threeparttable}
\end{table}

Table \ref{OOSPredict1} reports the out-of-sample results. Assessing the predictability of economic conditions, the full predictive regression with aggregate network connectedness yields significantly lower forecast errors for CFNAI at all forecast horizons and the indicator of NBER recession for 3 and 12 horizons. When both good and bad networks are added to the predictive regression, the statistics for the CFNAI and NBER recession errors improve, and we also improve our forecasts of industrial production. Looking at uncertainty indicators, we find that network connectedness improves the predictions of all uncertainty indicators across all forecast horizons significantly, with the only exception being macroeconomic average volatility.

Overall, both the in-sample and out-of-sample findings lead us to the conclusion that the information contained in the connectedness indexes, especially when decomposed, includes predictive information about a number of indicators of economic conditions and uncertainty, showing the usefulness of the forward-looking measures we have developed.

\section{Conclusion}    \label{concl}

The asymmetric network connectedness measures of fears were constructed to study the transmission of different shocks on fears extracted from the two sides of the stock options market in the U.S. financial network, as represented by the ten main U.S. financial institutions. The decomposed connectedness measures provide valuable forward-looking information, reflecting future investors' expectations about uncertainty.

Financial institutions play different roles as good/bad fear transmitters/receivers. From a systemic risk point of view, our new methodology provides a richer and more detailed picture of bank networks. For instance, we identify banks that are predominantly receivers of fear as well as those transmitting fear in the financial system. Being able to identify the more systemically important financial institutions can be helpful for preventing the spread of volatility and risk within the system, preparing financial institutions and policy makers to implement prudent operations in advance.

Having an ex ante monitoring tool for systemic risk is particularly useful for financial stability and market supervision. In addition, the asymmetric connectedness measures play an important role in signaling changes in future macroeconomic activity or uncertainty indicators. Our empirical analysis points out that there is significant predictive information in the good and bad fear network connectedness related to future macroeconomic activity or uncertainty, providing better predictive tools than the aggregate network.

\singlespace
\bibliography{SingleStock}
\bibliographystyle{chicago}

\end{document}